\documentclass{article}

\usepackage{arxiv}

\usepackage[utf8]{inputenc} % allow utf-8 input
\usepackage[T1]{fontenc}    % use 8-bit T1 fonts
\usepackage{hyperref}       % hyperlinks
\usepackage{url}            % simple URL typesetting
\usepackage{booktabs}       % professional-quality tables
\usepackage{amsfonts}       % blackboard math symbols
\usepackage{nicefrac}       % compact symbols for 1/2, etc.
\usepackage{microtype}      % microtypography
\usepackage{lipsum}		% Can be removed after putting your text content
\usepackage{graphicx}
\usepackage{natbib}
\usepackage{doi}
\usepackage{amsmath}

\title{The Glider Equation for Asymptotic Lenia}

\author{
    Hiroki Kojima\\
    the University of Tokyo, Japan\\
    Alternative Machine Inc., Japan\\ 
    \texttt{kojima@sacral.c.u-tokyo.ac.jp}\\
	\And
	Ivan Yevenko \\
    University of Waterloo, Canada\\
	\texttt{iyevenko@uwaterloo.ca}\\
	\AND
	Takashi Ikegami \\
    the University of Tokyo, Japan\\
    Alternative Machine Inc., Japan\\ 
	\texttt{ikeg@sacral.c.u-tokyo.ac.jp}\\
}

\renewcommand{\shorttitle}{The Glider Equation for Lenia}

\hypersetup{
pdftitle={The Glider Equation for Asymptotic Lenia},
pdfauthor={Kojima H., Yevenko I., Ikegami T.},
pdfkeywords={Lenia, Continous Celullar Automata, Neural Field, Glider},
}

\begin{document}
\maketitle

\begin{abstract}
Lenia is a continuous extension of Conway's Game of Life that exhibits rich pattern formations including self-propelling structures called gliders. In this paper, we focus on Asymptotic Lenia, a variant formulated as partial differential equations. By utilizing this mathematical formulation, we analytically derive the conditions for glider patterns, which we term the ``Glider Equation.'' We demonstrate that by using this equation as a loss function, gradient descent methods can successfully discover stable glider configurations. This approach enables the optimization of update rules to find novel gliders with specific properties, such as faster-moving variants. We also derive a velocity-free equation that characterizes gliders of any speed, expanding the search space for novel patterns. While many optimized patterns result in transient gliders that eventually destabilize, our approach effectively identifies diverse pattern formations that would be difficult to discover through traditional methods. Finally, we establish connections between Asymptotic Lenia and neural field models, highlighting mathematical relationships that bridge these systems and suggesting new directions for analyzing pattern formation in continuous dynamical systems.
\end{abstract}

% keywords can be removed

\section{Introduction}

Cellular Automata (CA) have been fundamental tools for studying emergent complexity from simple local rules. Conway's Game of Life is a classic example that demonstrates how rich behaviors including self-replication and universal computation can emerge from elementary rules \citep{gardner1970GoL}. However, traditional CA are discrete in both space and time, limiting their connection to natural phenomena that often exhibit continuous dynamics.

Lenia \citep{chan2019, chan2020} represents a significant advancement as a continuous generalization of cellular automata. Unlike the binary states and discrete updates of traditional CA, Lenia features continuous state values and smooth state transitions through the application of convolution kernels. This continuous framework allows for the emergence of diverse patterns with organic appearances, including self-propelling structures called ``gliders'' that resemble simple life forms. Recent work has further characterized the rich dynamics of Lenia and, for example, \citet{yevenko2024classifying} demonstrated the existence of fractal basins of attraction in Lenia's pattern space, highlighting the system's complex nonlinear behavior.

While the original Lenia cannot be directly described by differential equations due to its clipping operation \citep{tyrell2022step,kojima2023}, variants have been proposed to bridge this gap. Asymptotic Lenia \citep{kawaguchi2021introducing} reformulates the system as partial differential equations (PDEs), making it amenable to analysis using the established mathematical frameworks of continuous dynamical systems. Using this formulation, \citet{kojima2023} demonstrated that Asymptotic Lenia can be expressed as a reaction-diffusion system.

Previous approaches to discovering glider patterns in Lenia-like systems have primarily relied on manual exploration, random search, or evolutionary algorithms \citep{chan2020}. While these methods have successfully revealed interesting patterns, they lack the systematic approach that analytical formulations can provide. Recent work by \citet{hamon2024discovering} has explored the use of diversity search to discover sensorimotor agency in cellular automata, highlighting the richness of the pattern space that remains to be explored.

The present work focuses on the PDE formulation of Asymptotic Lenia, leveraging this mathematical representation to analytically derive the condition for glider patterns, which we term the ``Glider Equation.'' This analytical approach provides new insights into the formation and properties of gliders in continuous cellular automata systems. Furthermore, we demonstrate how this equation can be used as the basis for an optimization framework to discover and design glider patterns with specific properties.

The main contributions of this paper include the analytical derivation of the ``Glider Equation'' that characterizes self-propelling patterns in Asymptotic Lenia, along with a gradient-based optimization framework for systematically discovering diverse glider patterns. We further extend this framework to optimize update rules and discover patterns with specific properties, and develop a velocity-independent formulation of Glider Equation. Additionally, we identify and analyze connections between Asymptotic Lenia and neural field models, establishing links to established mathematical frameworks for studying pattern formation in continuous systems.

The paper is organized as follows: In Section 2, we review the mathematical formulation of Asymptotic Lenia. Section 3 presents the analytical derivation of the Glider Equation. Section 4 introduces our optimization-based approach to pattern discovery. Section 5 develops the velocity-free formulation of the Glider Equation. Finally, we discuss the implications of our findings, including connections to neural field theory, and conclude with future directions.

\section{Analytical Derivation of the Glider Equation}

\subsection{Mathematical Formulation of Asymptotic Lenia}

In two-dimensional space, the state evolution of Asymptotic Lenia is governed by:

\begin{equation}
    \frac{\partial}{\partial t}u(\mathbf{r}, t) = T(K * u(\mathbf{r}, t)) - u(\mathbf{r}, t)
\end{equation}

where $\mathbf{r} = (x, y)$ represents spatial coordinates.

The kernel $K$ typically has a radially symmetric form:
\begin{equation}
    K(\mathbf{r}) = \frac{1}{Z} k(|\mathbf{r}|/R)
\end{equation}
where $R$ is the kernel radius, $k$ is a kernel function, and $Z$ is a normalization factor ensuring that $\int K(\mathbf{r}) d\mathbf{r} = 1$. In practice, ring-shaped kernels are commonly used in Lenia systems, often constructed by stacking multiple concentric rings. In this work, we employ a three-ring kernel parameterization following the setup previously adopted in Asymptotic Lenia.

The target function $T$ is typically a ``bell-shaped'' function:
\begin{equation}
    T(x) = 2 \exp\left(-\frac{(x-\mu)^2}{2\sigma^2}\right) - 1
\end{equation}
where $\mu$ and $\sigma$ control the center and width of the bell curve, respectively. In Lenia systems, the parameter $\sigma$ is often set to small values, resulting in very sharp gaussian.

\subsection{Gliders as Traveling Wave Solutions}

In cellular automata, gliders represent localized patterns that move across the grid while maintaining their structure. In discrete systems like Conway's Game of Life, gliders typically undergo cyclic changes during movement. In contrast, in continuous cellular automata like Lenia, glider patterns that maintain their exact shape while translating across space are also observed.

In this paper, we focused on this class of glider. This condition can be mathematically formulated as follows. If a pattern moves with a constant velocity vector $\mathbf{v}$ without changing shape, then the value at $\mathbf{r}$ at time $t$ should be the same as the value at $\mathbf{r} - \mathbf{v}t$ at time 0. This can be expressed as:

\begin{equation}
    u(\mathbf{r},t) = u(\mathbf{r} - \mathbf{v}t, 0)
\end{equation}

This is precisely the mathematical definition of a traveling wave solution. We note that the PDE formulation of Asymptotic Lenia enables us to establish this direct connection between gliders and traveling waves, opening the door to applying the rich theoretical framework of traveling wave analysis. While traveling wave theory has been extensively developed in various contexts, its application to Lenia-like systems represents a novel direction, as we will discuss further in relation to neural field models.

By combining this traveling wave ansatz with the equations of Asymptotic Lenia, we can derive analytical conditions that any glider pattern must satisfy. Substituting the traveling wave ansatz into the Asymptotic Lenia equation requires computing the time derivative using the chain rule:

\begin{equation}
    \frac{\partial}{\partial t}u(\mathbf{r}, t) = -\mathbf{v} \cdot \nabla u(\mathbf{r} - \mathbf{v}t, 0)
\end{equation}

Substituting back into the Asymptotic Lenia equation and defining $\mathbf{r}' = \mathbf{r} - \mathbf{v}t$ for simplicity:

\begin{equation}
    -\mathbf{v} \cdot \nabla u(\mathbf{r}') = T(K * u(\mathbf{r}')) - u(\mathbf{r}')
\end{equation}

Rearranging and dropping the prime notation for clarity (as the equation must hold for all spatial positions), we obtain:

\begin{equation}
    u(\mathbf{r}) - \mathbf{v} \cdot \nabla u(\mathbf{r}) - T(K * u(\mathbf{r})) = 0
\end{equation}

This equation, which we term the ``Glider Equation,'' represents the analytical condition for a pattern to maintain its shape while moving with velocity $\mathbf{v}$. The derivation marks a significant theoretical achievement: for the first time, we have an analytical characterization of glider patterns in continuous cellular automata, shifting from purely empirical observation to mathematical understanding. Any spatial pattern $u(\mathbf{r})$ that satisfies this equation will propagate coherently with velocity $\mathbf{v}$ without deformation.

\section{Optimization-Based Glider Discovery}

\subsection{Reformulation as an Optimization Problem}

While the analytical derivation of the Glider Equation provides a theoretical foundation for understanding glider patterns in Asymptotic Lenia, finding closed-form solutions for this nonlinear equation is generally challenging. To bridge this gap between theory and practical application, we reformulate the problem as an optimization task.

Specifically, we seek spatial patterns $u(\mathbf{r})$ that minimize the following functional:

\begin{equation}
    F[u] = \int \left|u(\mathbf{r}) - \mathbf{v} \cdot \nabla u(\mathbf{r}) - T(K * u(\mathbf{r}))\right|^2 d\mathbf{r}
\end{equation}

This loss function represents the squared deviation from the Glider Equation across the entire spatial domain. It equals zero if and only if $u(\mathbf{r})$ is an exact solution to the Glider Equation. Therefore, by minimizing this functional through gradient-based optimization, we can systematically search for patterns that maintain their shape while moving at a specified velocity $\mathbf{v}$—the defining characteristic of gliders.

\subsection{Implementation and Experimental Setup}

We approach this minimization problem using gradient descent methods. Starting with an initial pattern $u_0$, we iteratively update:

\begin{equation}
    u_{n+1} = u_n - \eta \frac{\delta F}{\delta u}
\end{equation}

where $\eta$ is a learning rate and $\frac{\delta F}{\delta u}$ is the functional derivative of $F$ with respect to $u$.

We implemented this optimization approach using PyTorch. The system was discretized on a grid of size 144×144 pixels with periodic boundary conditions. Spatial gradients were computed using centered periodic differences. For convolution operations, we utilized the Fast Fourier Transform (FFT) to improve computational efficiency.
For most experiments, we used the Adam optimizer with learning rates ranging from $10^{-2}$ to $10^{-4}$ depending on the parameter type, and employed a step learning rate scheduler that decreased the learning rate by half every 1000 iterations. Typical optimization runs consisted of 5000 iterations.

For the initial pattern $u_0$, we primarily used Gaussian distributions with various width to promote the emergence of localized patterns rather than dispersed or uniform states. The kernel function $K$ was implemented as a weighted sum of three concentric ring basis functions, following earlier work on Lenia systems.

\subsection{Pattern Optimization with Fixed Rules}

We first evaluated our approach in a controlled setting where the existence of gliders was already established. This allowed us to validate whether our optimization framework could successfully recover known glider patterns from simple initial conditions. For these experiments, we fixed the parameters of the target function $T$ ($\mu=0.21$, $\sigma=0.018$) and the kernel function $K$ (with ring weights $[5/6, 7/12, 1]$), which are known to support glider patterns. The target velocity was initially set to 3.4 pixels per time unit, corresponding to the approximate speed of known gliders in this parameter regime.

\subsubsection{Emergence of Glider Patterns Through Gradient Descent}

\begin{figure}
    \centering
    \includegraphics[width=\linewidth]{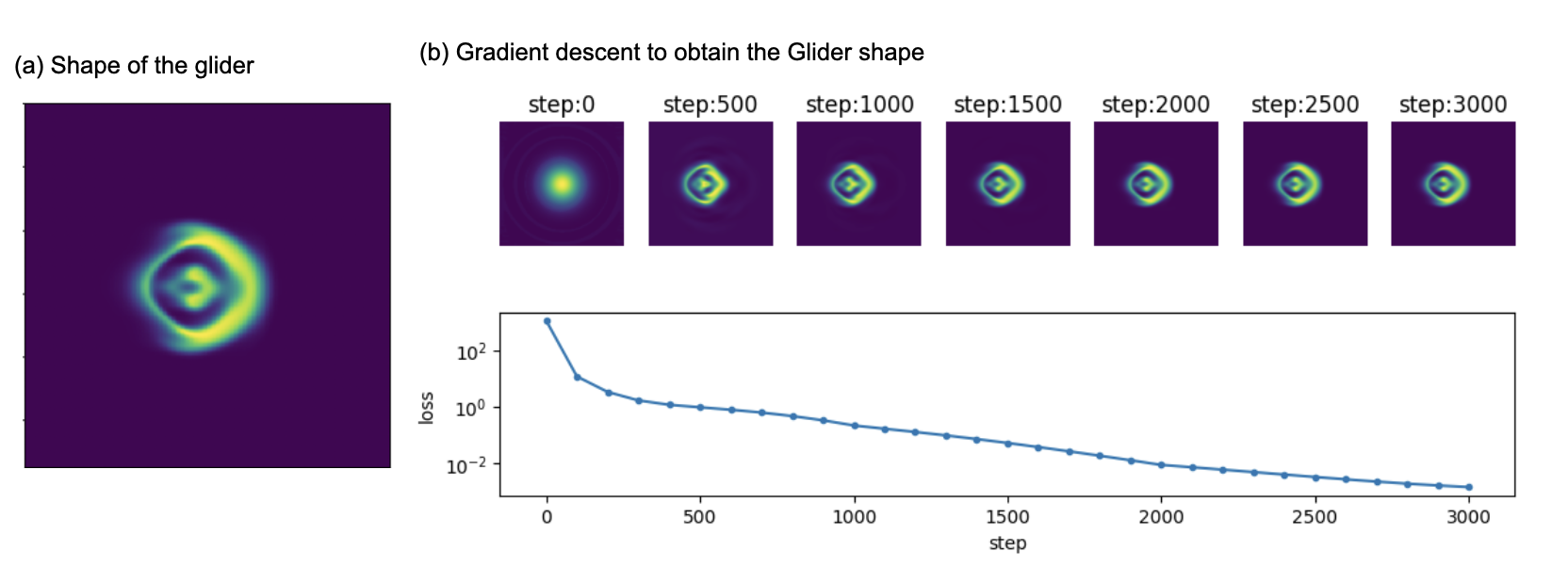}
    \caption{
        Obtaining glider shape through gradient descent optimization. (a) The known glider pattern in the update rule used here. (b) Top: The loss function decreases during optimization, indicating convergence toward a solution of the glider equation. Bottom: Initial Gaussian pattern evolves through the optimization process (from left to right) toward a stable glider shape.
    }
    \label{fig1}
\end{figure}

Figure~\ref{fig1} shows the results of applying gradient descent to find a glider pattern with fixed update rules. Starting from a Gaussian distribution (with $\sigma=15$), the optimization process progressively transforms the initial pattern into a structure closely resembling known gliders in this parameter setup. 

The loss function exhibits a rapid decrease during the early iterations, indicating that the pattern quickly approaches a solution to the Glider Equation. As optimization progresses, the morphological features of the gliders appears, and the final pattern obtained after optimization closely matches known gliders with these parameters.

Here, we noted that in this experiment, the loss function typically decreased to around 0.1 but did not reach zero. From the perspective of the Glider Equation, a perfect glider pattern should yield a loss value of zero. This discrepancy suggests that either the optimization process failed to find the global minimum, or the specified conditions do not permit an exact solution. We suspected that the preset velocity might not precisely match the natural velocity of the glider for these parameters.

To investigate this hypothesis further, we conducted additional experiments where we also updated the velocity vector during gradient descent rather than keeping it fixed. With this modification, we observed that the loss decreased to values as low as $10^{-5}$, while the velocity converged to a more accurate value. This confirms our hypothesis that the pre-specified velocity was slightly misaligned with the natural velocity of the optimal glider solution for these parameters.

\subsubsection{Dependency on Initial Conditions}

The results of our gradient descent optimization depend on both the initial pattern and the target velocity specified. Different initializations can lead to qualitatively different final patterns, even when the update rules remain fixed.

\begin{figure}
    \centering
    \includegraphics[width=\linewidth]{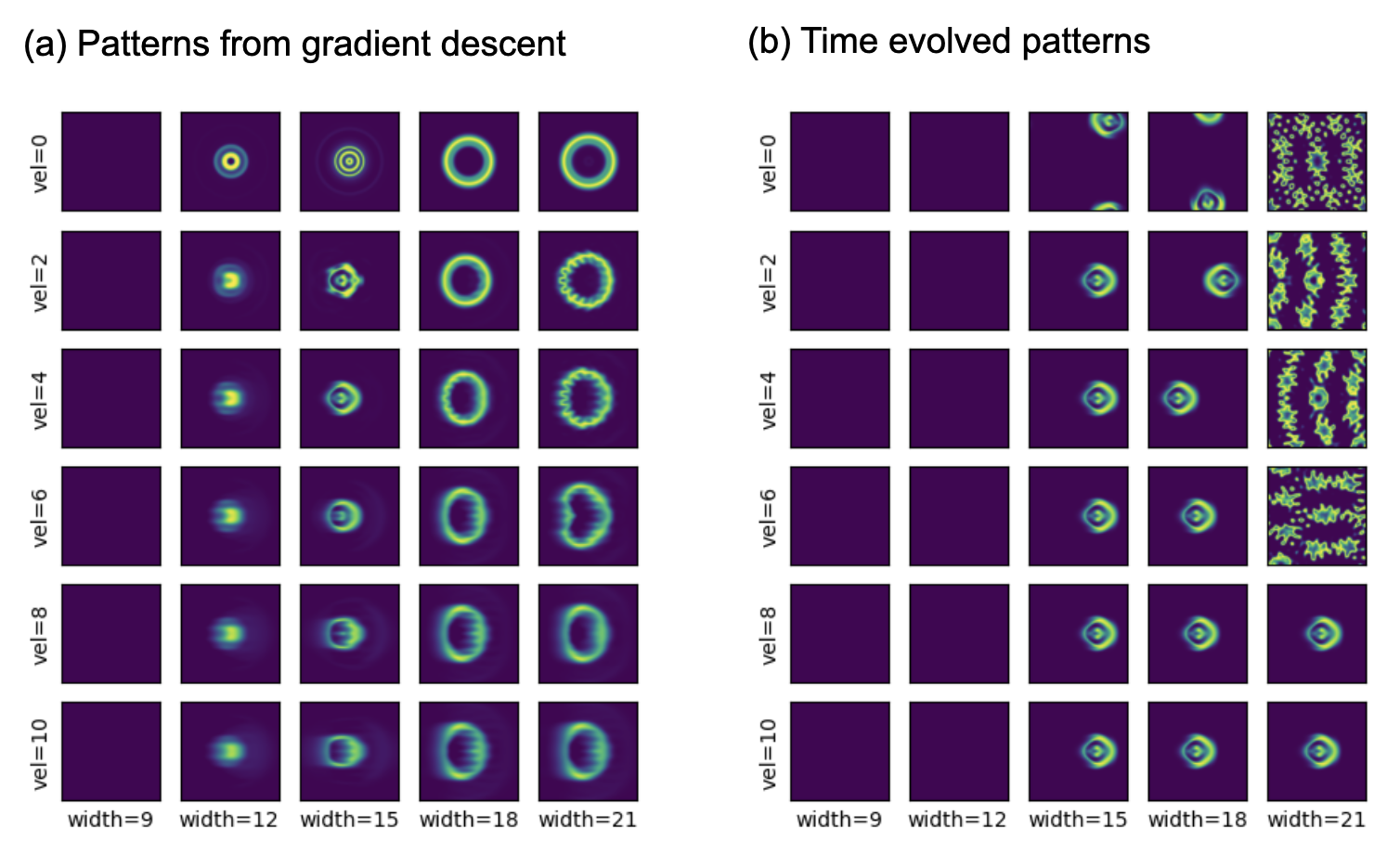}
    \caption{
        Dependency of optimization results on initial conditions. (a) Different initial patterns  lead to different glider shapes after optimization. (b) When these optimized patterns are time-evolved in Asymptotic Lenia, they exhibit distinctive motion behaviors. 
    }
    \label{fig2}
\end{figure}

Figure~\ref{fig2}(a) shows how these initial conditions affected the pattern obtained from the optimization. Especially, varying the width of the initial Gaussian pattern leads to convergence to different glider structures, despite using the same target velocity and update rules. This indicates that the solution space of the Glider Equation contains multiple local minima, each does not necessarily corresponds to stable glider. The optimization process tends to converge to the nearest local minimum from the starting point, and the barriers between these minima appear sufficiently high to prevent transitions between different glider configurations during the gradient descent.

Figure~\ref{fig2}(b) shows the time evolution of these optimized patterns when simulated in Asymptotic Lenia for 140 time steps (equivalent to 14 time units with $\Delta t = 0.1$). The patterns exhibit a variety of behaviors: some maintain their glider characteristics, others dissipate entirely, and some expand to fill the entire domain. This suggests that, though the optimized patterns have variations depending on the initial conditions, stable patterns are limited and eventually converged to one of these patterns.

Interestingly, patterns optimized with a target velocity of zero (representing stationary solutions) initially appear radially symmetric as shown in Figure~\ref{fig2}(a). However, when these patterns are evolved over time, they spontaneously break symmetry and develop directional motion, transforming into gliders with a specific orientation as seen in Figure~\ref{fig2}(b). This spontaneous symmetry breaking reveals the complex nonlinear dynamics underlying these systems.

\subsection{Rule Optimization for Novel Gliders}

While the previous section focused on optimizing pattern structures for fixed update rules, our framework can be extended to simultaneously optimize the rules themselves. By including the parameters of the target function $T$ and kernel function $K$ in the optimization process, we can search for novel combinations of rules and patterns that can support different glider behavior.

\subsubsection{Accelerating Glider Motion}

Inspired by recent work on discovering fast-moving patterns in cellular automata \citep{hamon2024discovering}, we first applied our approach to the challenge of finding faster gliders. Starting from the parameters that generate standard gliders with velocities around 3.4 pixels per time unit, we set a target velocity of 6.0 to encourage the optimization process to discover faster configurations.

We explored three optimization strategies of increasing flexibility:
1. Optimizing only the pattern structure while keeping rules fixed
2. Optimizing both the pattern and the parameters of the target function $T$
3. Optimizing the pattern, target function, and the kernel weights simultaneously

\begin{figure}
    \centering
    \includegraphics[width=0.9\linewidth]{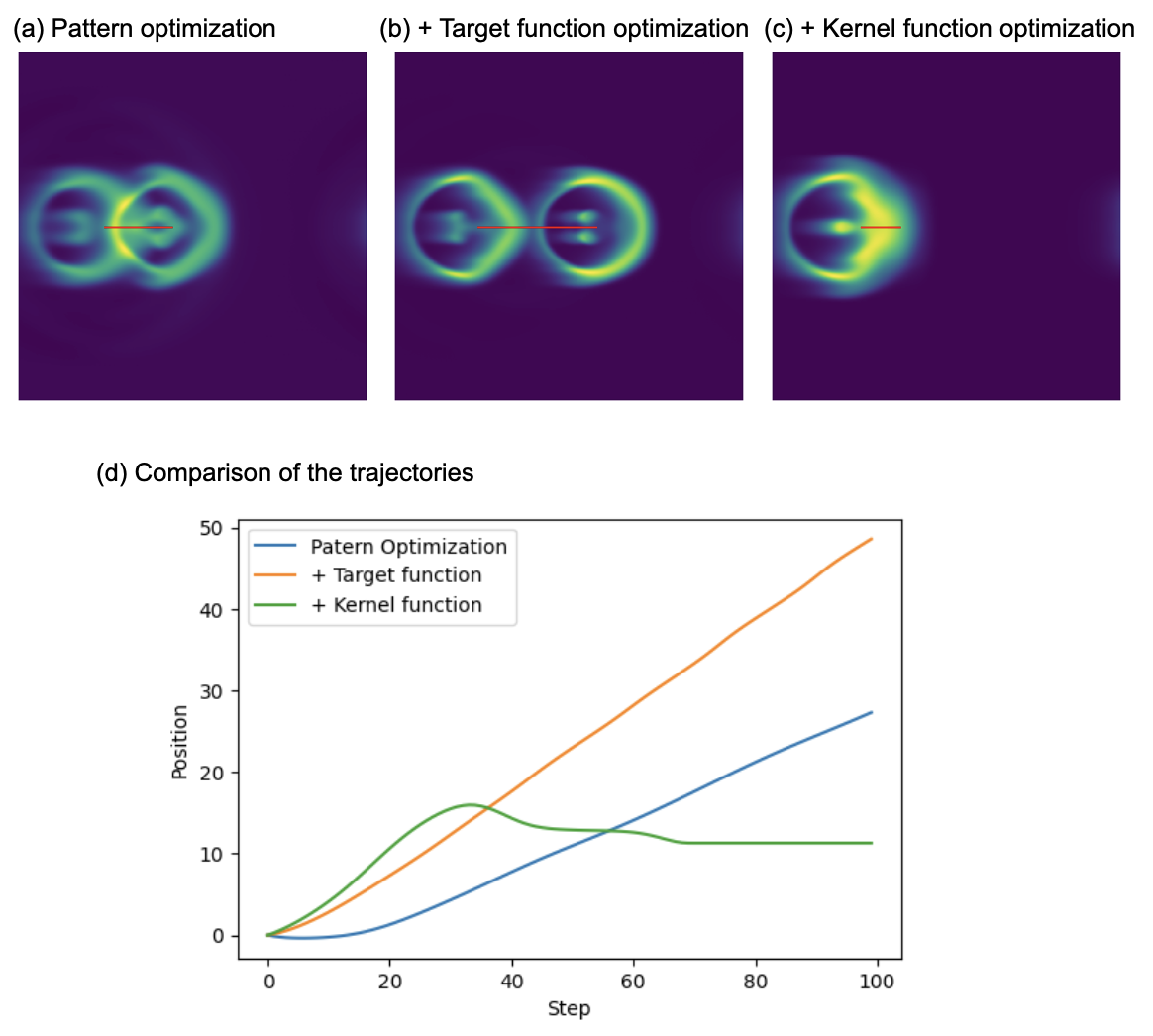}
    \caption{
        Gradient descent to create faster gliders by optimizing update rules. (a) Results when only updating the glider pattern while keeping rules fixed. (b) Results when optimizing the target function $T$ to facilitate faster motion. (c) Results when optimizing the kernel function $K$.
        In (a)-(c), the pattern after time evolution (40 steps) and the trajectory of the center of mass are overlaid. (d) The time series of the center of the mass of the pattern obtained from each condition.
    }
    \label{fig3}
\end{figure}

Figure~\ref{fig3} shows the results of these three approaches. When optimizing only the pattern (case 1), the solution converges back to the standard glider with its natural velocity, unable to achieve the target speed. When allowing the target function to vary (case 2), the optimization discovers a modified glider with a different morphology that moves somewhat faster, though still below the target velocity.

The most comprehensive optimization (case 3), which includes kernel weights, successfully produces patterns that approach the target velocity of 6.0 pixels per time unit. However, these accelerated gliders exhibit reduced stability. When simulated over extended periods, they eventually destabilize and dissipate. We observed that the gliders from case 2 also gradually destabilize over longer simulation periods, though they persist longer than those from case 3.

We observed that increasing the number of optimization iterations improved the longevity of these transient gliders before destabilization. This suggests that the optimization process can fine-tune patterns to achieve greater stability over time, though there may be fundamental speed limits for perfectly stable gliders in Asymptotic Lenia.

It is worth noting that the Glider Equation does not inherently assess the stability of the resulting patterns, only their conformity to the traveling wave condition. As a result, our optimization approach can discover transient patterns that satisfy the Glider Equation temporarily but eventually destabilize. This ability to systematically generate and study transient patterns represents an interesting feature of our approach, as these configurations can provide insights into the dynamics of pattern formation and dissolution in Asymptotic Lenia.

\subsubsection{Exploring Diverse Pattern Morphologies}

To more thoroughly explore the space of possible glider patterns, we conducted experiments with randomized initial conditions for the kernel weights rather than starting from known glider-supporting parameters. Combined with variations in target velocity and initial Gaussian widths, this approach enabled a broader search across the parameter space.

\begin{figure}
    \centering
    \includegraphics[width=0.9\linewidth]{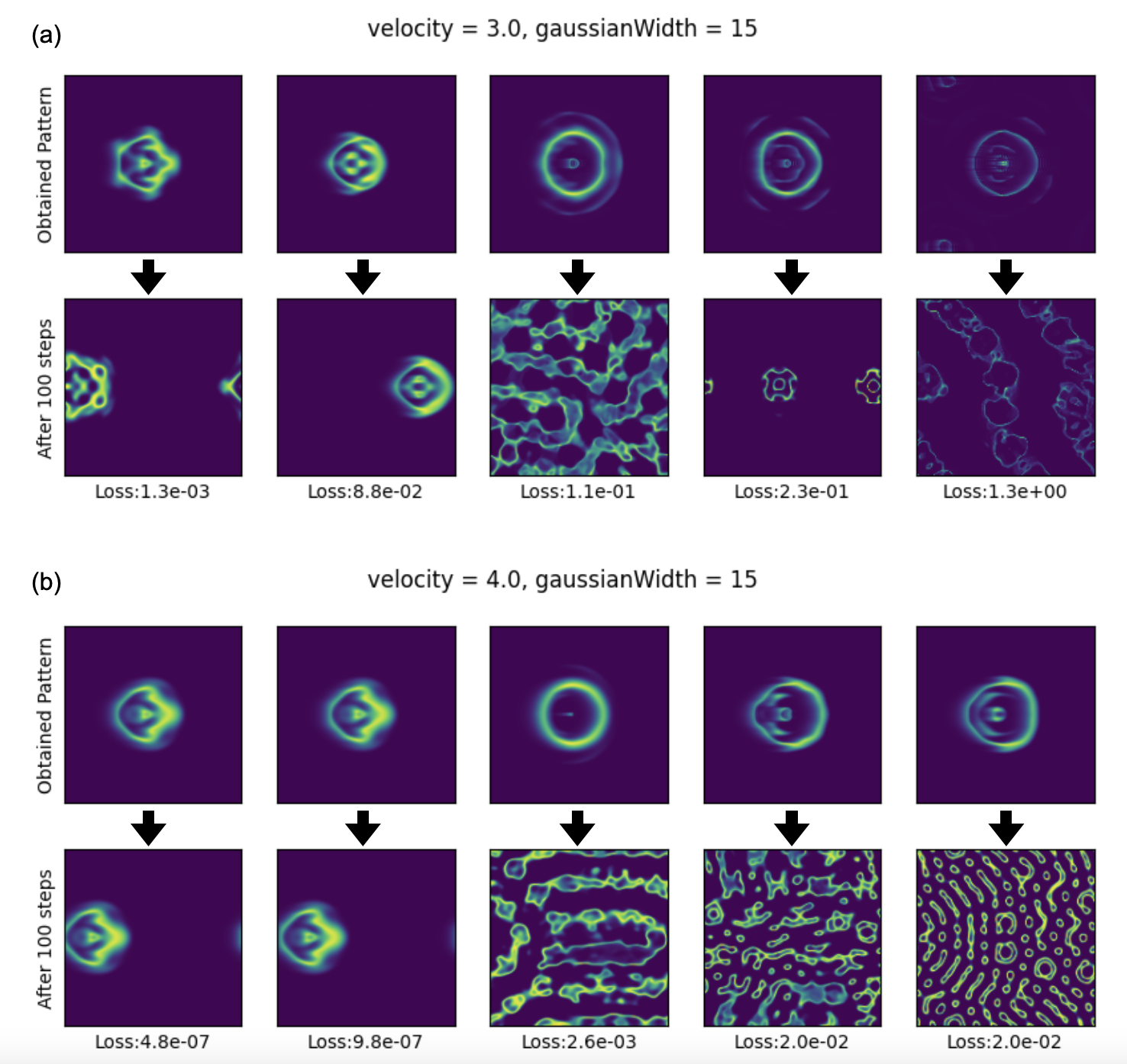}
    \caption{
        Discovery of diverse patterns through optimization with varied kernel weights. (a) Top row: Patterns obtained with a target velocity of 3.0 and initial Gaussian width of 15. Bottom row: The corresponding patterns after time evolution (100 steps). (b) Patterns obtained with a target velocity of 4.0 and initial Gaussian width of 15, and the corresponding patterns after time evolution (100 steps).
    }
    \label{fig4}
\end{figure}

For each condition (specific target velocity and Gaussian width), we generated 20 different patterns through our optimization process. We then time-evolved these patterns in Asymptotic Lenia. In Figure ~\ref{fig4}, we present the five patterns with the lowest loss values that maintained without dissipating after time evolution.

Figure ~\ref{fig4}(a) shows the results with a target velocity of 3.0 and an initial Gaussian width of 15. Among these patterns, the second from the left closely resembles the previously known gliders discussed in earlier sections, demonstrating that our optimization framework can rediscover established patterns while also revealing other interesting configurations.

In contrast, Figure ~\ref{fig4}(b) presents patterns obtained with a target velocity of 4.0 and the same initial Gaussian width. The first two patterns (from left) represent stable gliders with a morphology distinct from those in Figure ~\ref{fig4}(a). While they may appear visually similar (when horizontally reflected) to previously observed gliders, these patterns actually move rightward, essentially traveling in the ``opposite'' direction compared to the previous glider.

By jointly optimizing the kernel weights, target function parameters, and pattern structure, we discovered various patterns with different characteristics. This approach proved effective for exploring the various patterns of Asymptotic Lenia, including both stable gliders and interesting transient patterns that exhibit complex behaviors before eventually transforming or dissipating.

\section{Velocity-Free Formulation of the Glider Equation}

In the previous sections, we derived an equation for gliders with a pre-specified velocity vector and demonstrated its application in pattern optimization. We now return to theoretical foundations to develop a more general formulation that eliminates the need to specify velocity in advance. This velocity-free approach enables a more fundamental characterization of glider patterns and provides an alternative pathway for discovering self-propelling structures in Asymptotic Lenia.

\subsection{Theoretical Derivation}

To derive this generalized equation, we first define the residue obtained from the glider equation:

\begin{equation}
R(\mathbf{r}) := u(\mathbf{r}) - T(K * u(\mathbf{r})) = \mathbf{v} \cdot \nabla u(\mathbf{r})
\label{eq:residual}
\end{equation}

If we assume $u(\mathbf{r})$ is a glider, then second equality holds and this residue equals the directional derivative of $u$ along the velocity vector. We can use this relation to determine the velocity of a pattern without specifying it beforehand.

The spatial derivatives $\partial_x u$ and $\partial_y u$ span the two-dimensional subspace associated with translations (Goldstone modes). Taking the $L^2$ inner product of Equation~\ref{eq:residual} with each directional derivative (for $i \in \{x,y\}$):

\begin{equation}
\int R \, \partial_i u \, d^2r = v_x \int \partial_x u \, \partial_i u \, d^2r + v_y \int \partial_y u \, \partial_i u \, d^2r
\end{equation}

We define the following quantities:
\begin{align}
G_{ij} &:= \int (\partial_i u)(\partial_j u) \, d^2r\\
B_i &:= \int R \, \partial_i u \, d^2r
\end{align}

The two linear equations above can then be written compactly as:
\begin{equation}
G\mathbf{v} = \mathbf{B}
\end{equation}

or explicitly:
\begin{equation}
G_{xx}v_x + G_{xy}v_y = B_x, \quad G_{xy}v_x + G_{yy}v_y = B_y
\end{equation}

\subsection{Analytical Calculation of Glider Velocity}

Provided that the Gram matrix $G$ is nonsingular (i.e., $\partial_x u$ and $\partial_y u$ are linearly independent), the estimated velocity $\tilde{\mathbf{v}}$ is given by:

\begin{equation}
\tilde{\mathbf{v}} = G^{-1}\mathbf{B}, \quad \det G = G_{xx}G_{yy} - G_{xy}^2 \neq 0
\label{eq:velocity}
\end{equation}

Explicitly:
\begin{equation}
\tilde{v_x} = \frac{B_xG_{yy} - B_yG_{xy}}{\det G}, \quad \tilde{v_y} = \frac{B_yG_{xx} - B_xG_{xy}}{\det G}
\end{equation}

\begin{figure}
    \centering
    \includegraphics[width=\linewidth]{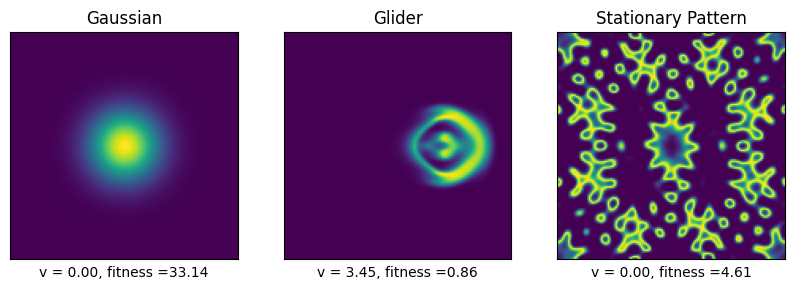}
    \caption{
        Calculation of velocity directly from pattern structure using the velocity-free formulation and velocity-free fitness values from three patterns, gaussian, glider and stationary pattern.
    }
    \label{fig5}
\end{figure}

This formula allows us to compute the velocity vector directly from the spatial pattern $u$ and the system dynamics, without resorting to numerical simulations or time-series analysis (Figure ~\ref{fig5}).

\subsection{Velocity-Free Optimization Framework}

Building on this theoretical foundation, we can develop an optimization framework that does not require specifying target velocities in advance. Inserting Equation~\ref{eq:velocity} back into Equation~\ref{eq:residual}, we obtain the error:

\begin{equation}
\varepsilon(\mathbf{r}) := R(\mathbf{r}) - \tilde{\mathbf{v}} \cdot \nabla u(\mathbf{r})
\end{equation}

We define a scalar fitness function as the $L^2$ norm of this error:

\begin{equation}
\mathcal{F}[u] = \|\varepsilon\|_2 = \left(\int \varepsilon^2 \, d^2r\right)^{1/2}
\label{eq:fitness}
\end{equation}

This fitness function vanishes if and only if $u$ is an exact traveling wave solution of the Asymptotic Lenia equations. Basically, this fitness function corresponds to inserting the estimated velocity $\tilde{\mathbf{v}}$ to the previous Glider equation. By minimizing $\mathcal{F}$, we can identify patterns that move with constant velocity without specifying their velocity in advance.

We confirmed that the fitness value of both the glider pattern and the stationary pattern were small compared to other patterns such as gaussian pattern (Figure ~\ref{fig5}).

\begin{figure}[ht]
    \centering
    \includegraphics[width=0.8\linewidth]{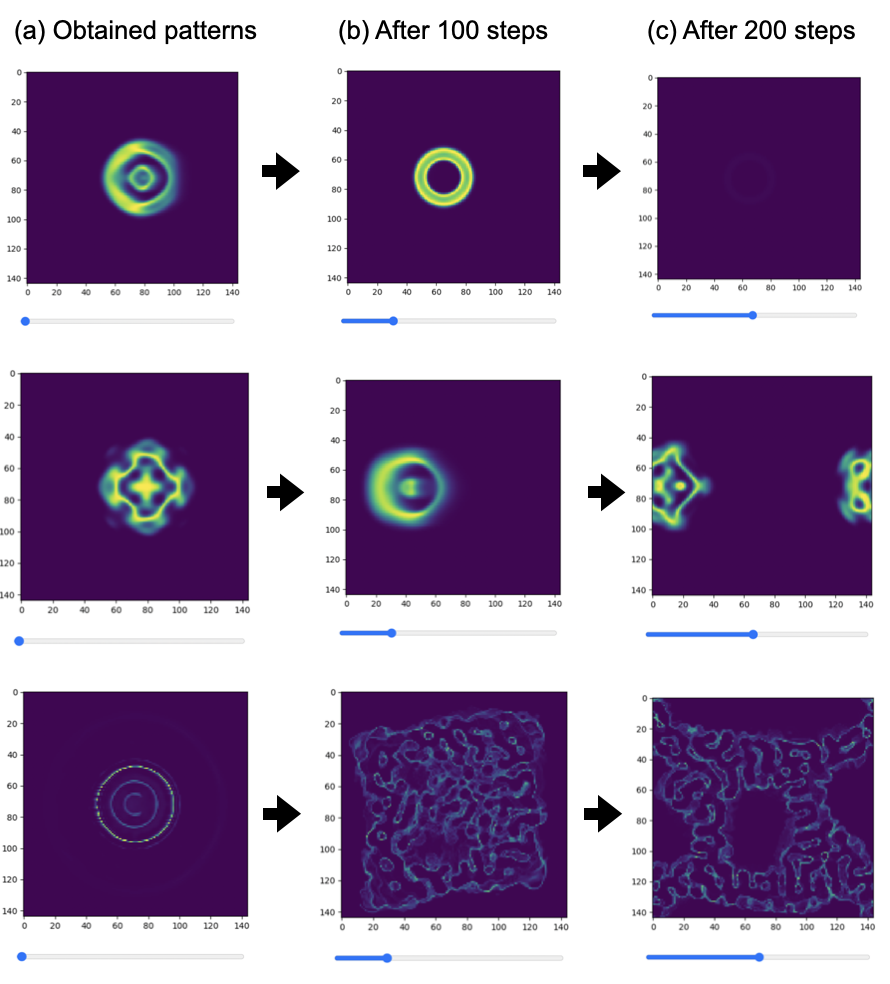}
    \caption{
        Temporal evolution of patterns discovered through velocity-free optimization. (a) Three distinct patterns obtained from random initial conditions using the velocity-free fitness function. (b) The same patterns after 100 time steps of evolution in Asymptotic Lenia. (c) After 200 time steps, showing diverse long-term behaviors including stable orbits, structured patterns, and complex space-filling dynamics.
    }
    \label{fig6}
\end{figure}

\subsection{Experimental Results with Velocity-Free Optimization}

While the velocity-free formulation is theoretically elegant, implementing it for pattern discovery requires addressing certain practical challenges. In our experiments, we found that the basic optimization of the fitness function $\mathcal{F}$ alone tends to converge toward trivial solutions.

\subsubsection{Symmetry Breaking and Pattern Diversity}

In practice, we found it necessary to add additional terms to the fitness function to break symmetry and guide the optimization toward non-trivial patterns. Without such terms, the optimization frequently converges to uniform or radially symmetric solutions that don't exhibit directional motion, despite satisfying the mathematical conditions.

To encourage the emergence of asymmetric, direction-specific patterns, we augmented the fitness function with a directional bias term:

\begin{equation}
\mathcal{F}_{\text{biased}}[u] = \|\varepsilon\|_2 + \lambda \exp\left(-\frac{v_x^2}{2\sigma^2}\right)
\end{equation}

where $\lambda$ controls the strength of the bias, $\sigma$ determines its width, and $v_x$ is the x-component of the calculated velocity. This Gaussian term penalizes patterns with zero or near-zero horizontal velocity, effectively breaking the radial symmetry and guiding the optimization toward patterns with clear directional motion. By tuning the parameters $\lambda$ and $\sigma$, we can adjust the pressure toward directional patterns while still allowing the optimization to find natural traveling wave solutions.

Figure~\ref{fig6} shows examples of patterns discovered through optimization of our enhanced velocity-free fitness function (Here, we set  $\lambda = 1.0$ and $\sigma = 0.1$). Some of these patterns overlap with those found using the velocity-specified approach from the previous section, but the velocity-free method offers greater flexibility by allowing the natural velocity to emerge from the optimization process itself.

This velocity-free approach complements our earlier method by providing an alternative pathway for pattern discovery. While the velocity-specified method allows targeted search for gliders with particular speeds, the velocity-free approach can be more effective for broad exploration of the pattern space without preconceptions about the expected motion characteristics.

\section{Discussion}

In this paper, we have derived analytical conditions for glider patterns in Asymptotic Lenia and demonstrated their application in pattern discovery. By formulating the Glider Equation based on the PDE representation of Asymptotic Lenia, we established a mathematical foundation for understanding self-propelling structures. We then showed how this equation can be reformulated as optimization problems, both with specified velocities and in a velocity-free formulation, enabling systematic discovery of diverse glider patterns. Our approach has revealed not only stable gliders but also transient patterns that temporarily satisfy the glider conditions before transforming or dissipating.

\subsection{Connection to Neural Field Models}

An important observation from our work is that Asymptotic Lenia can be categorized as a variant of neural field models proposed by Amari \citep{amari1977} and Wilson-Cowan \citep{wilson1972}, which describe the dynamics of neural activity across continuous neural tissues.

The standard neural field equation takes the form:
\begin{equation}
    \tau\frac{\partial u(\mathbf{r},t)}{\partial t} = -u(\mathbf{r},t) + \int w(|\mathbf{r}-\mathbf{r}'|)f(u(\mathbf{r}',t))d\mathbf{r}'
\end{equation}
where $\tau$ is a time constant, $w$ is a connectivity kernel, and $f$ is an activation function.

Comparing with the Asymptotic Lenia equation:
\begin{equation}
    \frac{\partial u(\mathbf{r},t)}{\partial t} = T(K * u(\mathbf{r},t)) - u(\mathbf{r},t)
\end{equation}
We can see a direct correspondence if we identify $K$ with $w$ and $T$ with $f$.

Pattern formation in neural field models has been extensively studied \citep{coombes2005waves,bressloff2011spatiotemporal,wilson2021evolution}, with traveling wave analysis dating back to Amari's early work \citep{amari1977}. While the mathematical derivation of our glider equations parallels this analysis, there are important structural differences between Asymptotic Lenia and typical neural field formulations that may account for the unique patterns observed in Lenia systems.

The most significant difference lies in the order of operations. In Asymptotic Lenia, the nonlinear function $T$ is applied after convolution ($T(K*u)$), whereas in standard neural field models, the nonlinear function is often applied before convolution ($K*f(u)$). Although the Lenia configuration is equivalent to what Wilson-Cowan termed the ``activity-based'' model \citep{wilson1972}, the convolution-first approach has received less mathematical attention compared to the activation-first formulation, which is more amenable to Fourier analysis.

Furthermore, the kernel characteristics differ significantly between these systems. Lenia typically employs kernels with $K(0)=0$ (center-surround or ring-shaped), which can be interpreted in neural field terminology as primarily extrinsic connections without intrinsic connections \citep{pinotsis2013dynamic}. Such kernel structures are less common in traditional neural field studies.

Additionally, the bell-shaped target function with narrow peaks used in Lenia differs substantially from the sigmoidal functions common in neural field models. This difference appears critical for the rich pattern formation observed in Lenia. We conjecture that the $T(K*u)$ configuration with its sharp activation profile enhances synergistic interactions \citep{mediano2022greater} between different regions of the pattern, serving as a driving mechanism for the emergence of complex structures.

To our knowledge, the analysis of traveling wave solutions in systems precisely equivalent to Asymptotic Lenia remains relatively unexplored in the neural field literature, making our analytical approach a novel contribution that bridges these related but distinct mathematical frameworks.

\subsection{The Value of Analytical Approaches in CA Research}

Our analytical derivation of the Glider Equation represents a significant advancement in the theoretical understanding of continuous cellular automata. By formulating the conditions for self-propelling patterns in mathematical terms, we provide a framework that complements the simulation-based approaches traditionally used in CA research.

This analytical approach offers several advantages: it provides insight into the fundamental mechanisms underlying pattern formation and self-propulsion; it enables systematic exploration of the pattern space through optimization techniques; and it establishes connections to other fields of mathematics and physics, particularly continuous dynamical systems.

From a computational perspective, our gradient-based approach offers potential efficiency advantages compared to traditional search methods. Our implementation requires evaluating only a single pattern state at a time rather than simulating system evolution over multiple time steps. However, we observe that the optimization landscape contains many local minima, which can limit the effectiveness of pure gradient descent methods.

This suggests that hybrid approaches combining our analytical framework with genetic algorithms or other global optimization methods might be particularly effective. For instance, using our velocity-free fitness function within a genetic algorithm framework could potentially leverage the strengths of both approaches—the theoretical guidance from our analytical conditions and the global search capabilities of evolutionary methods.

\subsection{The Significance of Transient Patterns}

The discovery of transient patterns—configurations that temporarily exhibit organized behavior before transforming or dissipating—is a noteworthy outcome of our approach. The accelerated gliders shown in Figure~\ref{fig3} and several patterns in the middle row of Figure~\ref{fig6} exemplify such transient formations, which satisfy the Glider Equation over short timescales but lack long-term stability.

These transient patterns represent an interesting class of solutions that has received little attention in previous research, partly because they are inherently difficult to encounter through conventional methods. Standard simulation approaches may not readily reveal these patterns, as they require specific initial conditions. Our observations of the accelerated pattern in Figure~\ref{fig3}, where increased optimization iterations yielded longer-lasting transient gliders, suggest that the fine-tuning of initial configurations plays a role in their emergence—sharing some conceptual similarities with ``Garden of Eden'' configurations in classical cellular automata theory.

Our gradient-based optimization approach, guided by the Glider Equation, provides an effective method to identify these transient configurations that may be difficult to discover through pure random search methods. This approach offers a systematic technique that expands our ability to explore specific regions of the pattern space, complementing other search strategies.

Studying these transient patterns may contribute to our understanding of the stability landscape of Lenia-like systems and the mechanisms by which patterns transform over time. While their practical significance remains an open question, they represent an interesting aspect of the rich dynamical behavior exhibited by continuous cellular automata.

\section{Conclusion}

Our derivation of the Glider Equation provides new insights into the mechanisms underlying persistent complex patterns in Lenia-type systems. By analytically characterizing the conditions for self-propelling structures, we have developed a theoretical framework that complements simulation-based approaches to studying continuous cellular automata.

The optimization-based implementation of this framework has enabled the discovery of diverse glider patterns, including both stable configurations and transient structures that would be difficult to find through random search methods. The velocity-free formulation further extends this approach, allowing patterns and their natural velocities to emerge simultaneously from the optimization process.

The connection we have established between Asymptotic Lenia and neural field models bridges two previously separate domains of pattern formation theories, opening new avenues for cross-fertilization between these fields. This integration enhances our understanding of emergent complexity in systems at the intersection of discrete and continuous dynamics.

By providing a mathematical foundation for understanding glider patterns in continuous cellular automata, this work contributes to the broader goal of connecting emergent computational phenomena to fundamental principles of dynamical systems. Future work can build on this foundation to explore stability conditions, pattern interactions, and deeper connections to other mathematical frameworks for understanding self-organization and emergence.

\section{Code Availability}

The code used in this study is available at  \url{https://github.com/hkjm/glider-equation-code/}.

\section{Acknowledgments}
This work was partially supported by the JSPS Kakenhi Grant-in-Aid [21H04885,24H00707].

\bibliographystyle{apalike}
\bibliography{references}  %%% Uncomment this line and comment out the ``thebibliography'' section below to use the external .bib file (using bibtex) .

\begin{thebibliography}{}

\bibitem[Amari, 1977]{amari1977}
Amari, S.-i. (1977).
\newblock Dynamics of pattern formation in lateral-inhibition type neural fields.
\newblock {\em Biological cybernetics}, 27(2):77--87.

\bibitem[Bressloff, 2011]{bressloff2011spatiotemporal}
Bressloff, P.~C. (2011).
\newblock Spatiotemporal dynamics of continuum neural fields.
\newblock {\em Journal of Physics A: Mathematical and Theoretical}, 45(3):033001.

\bibitem[Chan, 2019]{chan2019}
Chan, B. W.-C. (2019).
\newblock Lenia: Biology of artificial life.
\newblock {\em Complex Systems}, 28(3):251--286.

\bibitem[Chan, 2020]{chan2020}
Chan, B. W.-C. (2020).
\newblock Lenia and expanded universe.
\newblock In {\em Artificial Life Conference Proceedings 32}, pages 221--229. MIT Press One Rogers Street, Cambridge, MA 02142-1209, USA journals-info~….

\bibitem[Coombes, 2005]{coombes2005waves}
Coombes, S. (2005).
\newblock Waves, bumps, and patterns in neural field theories.
\newblock {\em Biological cybernetics}, 93:91--108.

\bibitem[Davis and Bongard, 2022]{tyrell2022step}
Davis, Q.~T. and Bongard, J. (2022).
\newblock Step size is a consequential parameter in continuous cellular automata.
\newblock In {\em ALIFE 2022: The 2022 Conference on Artificial Life}, Artificial Life Conference Proceedings, page~43.

\bibitem[Gardner, 1970]{gardner1970GoL}
Gardner, M. (1970).
\newblock Mathematical games : the fantastic combinations of john conway's new solitaire game 'life'.
\newblock {\em Scientific American}, 223:120--123.

\bibitem[Hamon et~al., 2024]{hamon2024discovering}
Hamon, G., Etcheverry, M., Chan, B. W.-C., Moulin-Frier, C., and Oudeyer, P.-Y. (2024).
\newblock Discovering sensorimotor agency in cellular automata using diversity search.
\newblock {\em arXiv preprint arXiv:2402.10236}.

\bibitem[Kawaguchi et~al., 2021]{kawaguchi2021introducing}
Kawaguchi, T., Suzuki, R., Arita, T., and Chan, B. (2021).
\newblock Introducing asymptotics to the state-updating rule in lenia.
\newblock In {\em Artificial Life Conference Proceedings 33}, volume 2021, page~91. MIT Press One Rogers Street, Cambridge, MA 02142-1209, USA journals-info~….

\bibitem[Kojima and Ikegami, 2023]{kojima2023}
Kojima, H. and Ikegami, T. (2023).
\newblock Implementation of lenia as a reaction-diffusion system.
\newblock In {\em ALIFE 2023: Ghost in the Machine: Proceedings of the 2023 Artificial Life Conference}. MIT Press.

\bibitem[Mediano et~al., 2022]{mediano2022greater}
Mediano, P.~A., Rosas, F.~E., Luppi, A.~I., Jensen, H.~J., Seth, A.~K., Barrett, A.~B., Carhart-Harris, R.~L., and Bor, D. (2022).
\newblock Greater than the parts: a review of the information decomposition approach to causal emergence.
\newblock {\em Philosophical Transactions of the Royal Society A}, 380(2227):20210246.

\bibitem[Pinotsis et~al., 2013]{pinotsis2013dynamic}
Pinotsis, D.~A., Schwarzkopf, D.~S., Litvak, V., Rees, G., Barnes, G., and Friston, K.~J. (2013).
\newblock Dynamic causal modelling of lateral interactions in the visual cortex.
\newblock {\em Neuroimage}, 66:563--576.

\bibitem[Wilson and Cowan, 1972]{wilson1972}
Wilson, H.~R. and Cowan, J.~D. (1972).
\newblock Excitatory and inhibitory interactions in localized populations of model neurons.
\newblock {\em Biophysical journal}, 12(1):1--24.

\bibitem[Wilson and Cowan, 2021]{wilson2021evolution}
Wilson, H.~R. and Cowan, J.~D. (2021).
\newblock Evolution of the wilson--cowan equations.
\newblock {\em Biological cybernetics}, 115(6):643--653.

\bibitem[Yevenko, 2024]{yevenko2024classifying}
Yevenko, I. (2024).
\newblock Classifying the fractal parameter space of the lenia orbium.
\newblock In {\em Artificial Life Conference Proceedings 36}, volume 2024, page~14. MIT Press One Rogers Street, Cambridge, MA 02142-1209, USA journals-info~….

\end{thebibliography}





% \end{thebibliography}

\end{document}